\documentclass[aps,pra,twocolumn,notitlepage,superscriptaddress,floatfix]{revtex4-1}

\usepackage{graphicx}
\usepackage{dcolumn}
\usepackage{bm}
\usepackage{amssymb}
\usepackage{microtype}
\usepackage{xfrac}
\usepackage{makecell}
\usepackage{array}
\usepackage[version=4]{mhchem}
\usepackage{gensymb}
\usepackage{multirow}
\usepackage{physics}
\usepackage[colorlinks=true]{hyperref}
\usepackage{physics}
\usepackage{bm}

\DeclareMathOperator{\sgn}{sgn}

\begin{document}

\title{Strain tuning the magnetic and transport properties of Mn$_3$Ge}
\author{Sayak Dasgupta}
\affiliation{Department of Physics and Astronomy $\&$ Stewart Blusson Quantum Matter Institute, University of British Columbia, Vancouver,  British Columbia V6T 1Z1, Canada}
\affiliation{Institute for Solid State Physics, University of Tokyo, Kashiwa 277-8581, Japan}

\begin{abstract}
The kagome lattice antiferromagnet Mn$_3$Ge has a local hexagonal symmetry with a 120$^\circ$ ordered ground state. This non-colinear ground state engenders a strong anomalous Hall response. The main goal of this work is to understand the effect of strain on this response. We derive an effective model for the Hall vector which functions as our order parameter. Using this model we show how both intra and inter planar strains can be used to switch the sign of the Hall response at a constant magnetic field. Further, we also investigate the effect of this strain on the spin wave band gaps and show that strain can be used to effectively manipulate them, which can be used to tune the magnon responses in this system.
\end{abstract}

\maketitle

Antiferromagnets hold a promise for a faster spintronics platform. The spin wave dynamics of an antiferromagnetic system is controlled by an energy scale proportional to $J$, where $J$ is the effective antiferromagnetic exchange. This translates to a frequency scale of a few THz. Antiferromagnets offer another significant advantage over ferromagnetic devices---the absence of stray fields from a net magnetization. This is particularly important in device design, where one wants individual memory components to be isolated from one another \cite{RMPYaroslav2018,Jungwirth2016}. A local density of magnetization is rendered energetically costly by the Heisenberg exchange. An effective field theory is expressed in terms of soft modes, which are spin configurations with vanishing net spin density. The magnetization density follows the soft mode dynamics and renders an inertial mass to the soft modes \cite{tveten14,dasgupta17}.

The family of non-collinear antiferromagnetic compounds Mn$_3$X (X = Ge, Sn $\cdots$) form a remarkable group of magnetic semimetals. All of them have a scaffold of Mn atoms arranged on a kagome lattice with a staggered AB stacking along the c-axis. The effective moments on the Mn sites are fairly large ($\simeq 1.78 \mu_B$) and are arranged in a 120$^\circ$ order in each triangle \cite{Nagamiya:1979,Tomiyoshi:1982a,Tomiyoshi:1982b}. Dzyaloshinskii-Moriya (DM) \cite{dzyaloshinsky58,moriya60} exchange, forces the spin ordering to be antichiral and coplanar with the kagome plane. The X site provides a local easy axis anisotropy which is much smaller than the exchange and DM interaction scales. The local anisotropy cants the moments in the kagome plane away from a perfect 120$^\circ$ order and creates a tiny dipolar moment per triangle, of the order of $10^{-3}\mu_B$ per Mn atom.

Despite this tiny ferromagnetic moment, both Mn$_3$Sn and Mn$_3$Ge show very large room temperature anomalous Hall conductivity $\sigma^H = 30 \Omega^{-1}$cm$^{-1}$ and $\sigma = 50-60 \Omega^{-1}$cm$^{-1}$ respectively \cite{Nakatsuji2015,nakatsujimn_3ge,nayak-hall-mn3ge}. This property is exclusive to the antichiral 120$^\circ$ ordered phase---owing to a large internal Berry curvature sourced from the Weyl points near the Fermi energy. This large Berry curvature in a compound at room temperature drives a growing interest in these materials, especially from the spintronics community, where it is seen as a potential device element \cite{Jungwirth2016b,Baltz2018,Smejkal2018}. 

The configurations of the local moments are best expressed in terms of the irreducible representation of the $D_{3h}$ point group. This decomposes the spin triangle into two sets of modes---$\alpha$ for distortions in the kagome plane and $\beta$ for distortions out of the plane \cite{dasgupta-mn3ge-2020,Zelenskiy-mn3ge-2021}. Of these, the mode which is vital to the transport properties of the compound is the singlet representing uniform rotations of all three spin in the $xy$ plane ($O_2$ symmetric), $\alpha_0$. Liu and Balents in \cite{Liu:2017} showed that the Hall vector characterizing the internal Berry phase is proportional to the amplitude of this mode. External control over this mode then enables us to manipulate the Berry curvature and hence the transport properties.

This was done using in plane shear strain on Mn$_3$Sn in Ikhlas $et~al$ \cite{ikhlas-future}. There a theoretical model was developed for an in plane strain which converts the $D_{3h}$ symmetry into a $C_2$ symmetry. This allows a $\pi$ rotation in the $\alpha_0$ mode at a constant magnetic field as strain is varied. An experiment was done to verify these predictions and determine the piezomagnetic coefficient. Our goal here is to look at the strain response in Mn$_3$Ge in presence of both and intraplanar and an interplanar shear strain. The advantage for the Ge compound stems from the fact that its antichiral phase persists at lower temperatures, whereas in Sn the helical phase takes over at 270K \cite{kiyohara2016,soh2020}. This has possible implications if we want to extend this phenomenology into a quantum regime. We do a detailed study of how misalignment and larger anisotropy might affect the switching parameters and also investigate how strain affects the spin wave bands in Mn$_3$X and the energy gaps at $\Gamma$ point.

\section{Magnetic Structure}

The magnetic structure of Mn$_3$Ge has been studied in detail through inelastic neutron scattering experiments \cite{Chen-mn3ge-2020,boothroyd-mn3ge-2020} and finite temperature Monte Carlo simulations \cite{chaudhary2021magnetism}. The ground state of this system is a $\mathbf{q} = 0$ ordered state where the local moments are arranged in a 120$^\circ$ order on each triangle. This is set by a nearest neighbour antiferromangetic Heisenberg exchange which is the dominant energy scale in the system. Secondary exchanges up to fourth nearest neighbour are required to fully model the spin wave spectra obtained from inelastic neutron scattering \cite{Chen-mn3ge-2020}. The arrangement of spins is antichiral---if one traverses the corners of the triangle in an anticlockwise manner the spins rotate clockwise. This chirality is set by a Dzyaloshinskii-Moriya (DM) interaction with an out of plane DM vector $\mathbf{D}_{ij} = D\hat{\mathbf{z}}$. The out of plane DM also forces the spins into the kagome plane with no discernible canting along the c-axis. In the kagome plane there is a second set of easy axes which point from the Mn sites to the centre of the triangle. The energetics is captured in the minimal model:
\begin{equation}
    \label{eq.minimal-H}
    \mathcal{H} = \sum_{ij} J_{ij} \mathbf{S}_i\cdot\mathbf{S}_j + D\sum_{<i,j>} \mathbf{\hat{z}}\cdot(\mathbf{S}_i \times \mathbf{S}_j) - \delta \sum_i (\mathbf{n}_i\cdot\mathbf{S}_i)^2
\end{equation}
Since the order is set by the Heisenberg exchange with $J \gg D \gg \delta$ we can take the order parameter of the system to be the orientation vector of the three spins per triangle locked into a triangle. This is an $SO(3)$ order parameter in a locally $D_{3h}$ symmetric environment and hence we expect three Goldstone modes with three corresponding sound velocities \cite{Chen-mn3ge-2020,dasgupta-mn3ge-2020}.

\begin{figure}[hbt]
		\centering
		\includegraphics[width = \columnwidth]{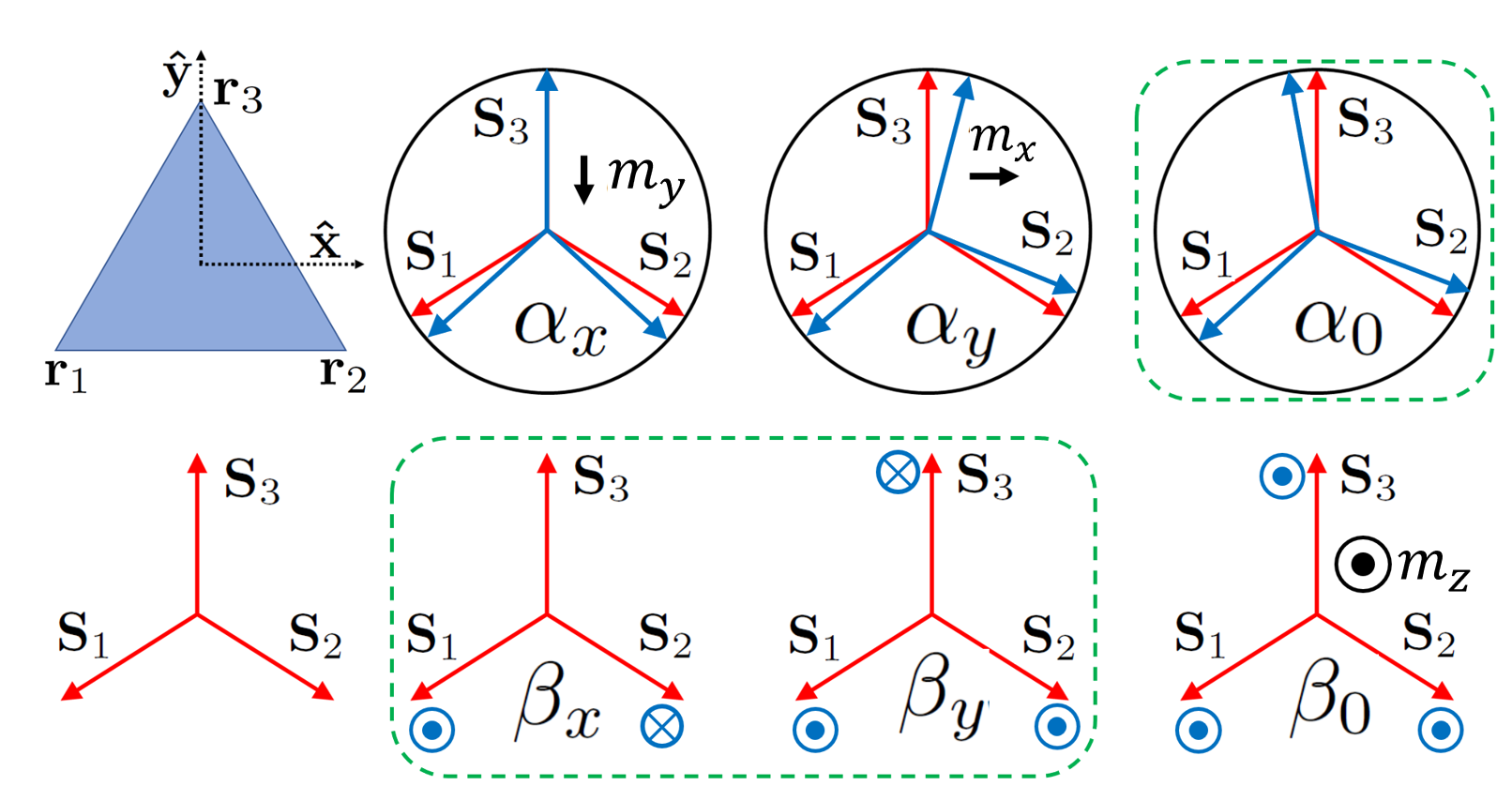}
		\caption{The geometry of the 120$^\circ$ state. The spins carry the same labels as the sites, spin $\mathbf{S}_i$ is at site $\mathbf{r}_i$. The normal mode decomposition is shown on the right. The red arrows indicated the ground state and the blue arrows the deviation represented by each normal mode. The magnetizations induced by the modes $\bm{\alpha}$ and $\beta_0$ are marked. The green dashed box includes the soft modes.}
		\label{fig.modes}
\end{figure}

\subsection{Normal modes}

From the irreducible representation of $D_{3h}$ we have three in plane modes $\alpha_0,\alpha_x,\alpha_y$, and three out of plane modes $\beta_0,\beta_x,\beta_y$. Of these $\bm{\alpha} = (\alpha_x,\alpha_y)$ and $\bm{\beta} = (\beta_x,\beta_y$) transform as doublets under the symmetry operations, see Fig.~\ref{fig.modes}. Modes $(\alpha_0,\bm{\beta})$ do not induce a net magnetization per triangle and are hence soft. The other three modes $(\bm{\alpha},\beta_0)$ induce an in-plane and an out of plane magnetization respectively and are hence penalized by the Heisenberg exchange \cite{dasgupta-mn3ge-2020}. 

We can think of these modes in terms of $-\beta_x,-\beta_y,\alpha_0$ as global rotation angles and $\alpha_x,\alpha_y$ and $\beta_0$ as the corresponding components of angular momentum along the lines of Mineev \cite{Mineev1996}. This also identifies the canonically conjugate pairs $\{-\beta_x,\alpha_x\}$, $\{-\beta_y,\alpha_y\}$, and $\{\alpha_0,\beta_0\}$. This identification allows us to write down the Berry phase term for the triangle:
\begin{equation}
    \mathcal{L}_B = \mathcal{S} (\Dot{\alpha}_0\beta_0 - \bm{\alpha}\cdot\Dot{\bm{\beta}}),
\end{equation}
where $\mathcal{S}$ is the spin density on a single sublattice. The leading order contribution to energetics is from the Heisenberg exchange interactions penalizing the hard modes. This results in a local theory of the form:
\begin{equation}
    \mathcal{L} = \mathcal{L}_B - \mathcal{U}(\bm{\alpha},\beta_0).
\end{equation}
If we restrict ourselves to quadratic order in fields in $\mathcal{U}$ we can integrate out the hard modes through their equations of motion to generate a kinetic term for the soft modes:
\begin{equation}
    \mathcal{K} = \frac{\rho_{\alpha}}{2} \dot{\alpha_{0}}^2 + \frac{\rho_{\beta_x}}{2} \dot{\beta}_x^2 + \frac{\rho_{\beta_y}}{2} \dot{\beta}_y^2,
\end{equation}
where $\rho$'s are functions of the exchanges. In the case of the nearest neighbour model we have $\rho_{\beta} = \rho_{\beta_x} = \rho_{\beta_y}$, but this in general does not hold when we consider further neighbour interactions \cite{dasgupta-mn3ge-2020}. The spin wave dispersion is then obtained by first calculating the leading order contribution from the soft modes to the potential energy, which are in the form of gradients of the modes. This results in the net Lagrangian density $\mathcal{L} = \mathcal{L}_{\text{kin}} - \mathcal{U}_g(\nabla\alpha_0,\nabla\beta_x,\nabla\beta_y)$, from which we can determine the spin wave dispersions.

Mn$_3$Ge is layered kagome system where the two layers are displace relative to each other such that the up triangles of one layer coincide with the down triangle of the layer above and below it. Each triangle in each layer is arranged in the $120^\circ$ anticlockwise state. The largest energy scale is the nearest neighbour exchange in a kagome layer. The bilayer nature doubles the number of modes from six to twelve---$(\alpha_{0}^A,\bm{\alpha}^A,\beta_{0}^A,\bm{\beta}^A)$ for the A layer and $(\alpha_{0}^B,\bm{\alpha}^B,\beta_{0}^B,\bm{\beta}^B)$ for the B layer. The better basis to work with is the symmetric and antisymmetric combination of these modes:
\begin{equation}
\label{eq.modes-interlayer}
\zeta = \frac{\zeta^A + \zeta^B}{\sqrt{2}}, ~~~ \Bar{\zeta} = \frac{\zeta^A - \zeta^B}{\sqrt{2}},
\end{equation}
where $\zeta$ stands for any of the $\alpha$ or $\beta$ modes. It might seem that this mode doubling results in a complicated theory, but it turns out that all the antisymmetric modes are rendered energetically unfavourable by the fourth neighbour exchange $J_4$ which is substantial in Mn$_3$Ge \cite{Chen-mn3ge-2020}. So in effect we only deal with the symmetric combinations. The detailed spin wave theory including the effect of anisotropies on the energy gaps was worked out in \cite{dasgupta-mn3ge-2020} and \cite{Zelenskiy-mn3ge-2021}. However, this theory does not take into account external straining which forms the focus of the current article.

\begin{figure}[t]
		\centering
		\includegraphics[width = .9\columnwidth]{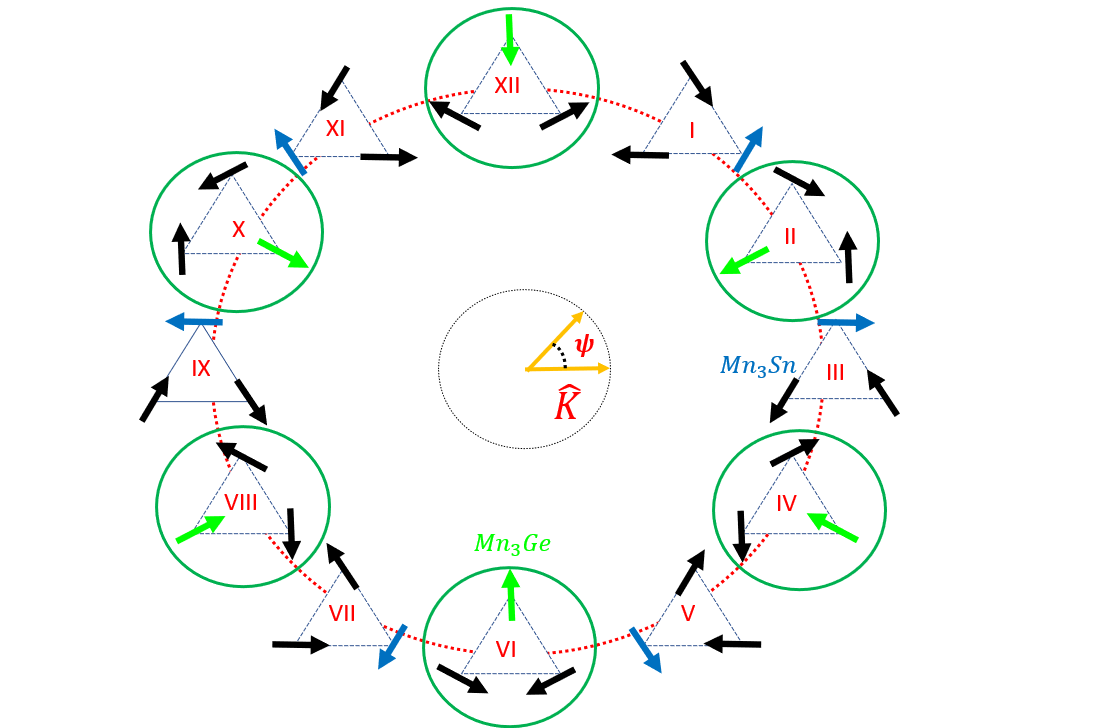}
		\caption{Mn$_3$X has a six-fold anisotropy, $C_6$, in the kagome plane rising from an easy axis, which points from each Mn site to the X site \cite{Liu:2017}. The two compounds Mn$_3$Sn and Ge differ in their easy axes directions by $\pi/2$. This results in six ground states for each compound which lie on the dial of a 12 hour clock as shown here. The odd hours are for Ge and the even hours are for Sn. Each state can be represented on a single triangular plaquette, with black arrows representing spins not aligned to the local easy axis and green representing spins pointing along the local easy axis for Mn$_3$Ge (blue for Mn$_3$Sn). The clock states can be represented by the vector $\mathbf{K} = (\cos\psi,\sin\psi)$ in the kagome plane. }
		\label{fig.geometry-modes}
\end{figure}

\section{effective theories in the kagome plane}
\label{sec.Effective_theories}
Let us consider a single kagome layer system. First we show that if the energetics is completely controlled by nearest neighbour exchange (J), DM interaction and easy plane anisotropy we get an effective theory where the $\alpha_0$ mode has six minima. Each of these represents a ground state of the system. In this minimal model the $\bm{\beta}$ modes have a higher energy than the $\alpha_0$ mode due to the DM interaction. 

The exchange interaction interaction expanded near the $\Gamma$ point to quadratic order in soft and hard modes:
\begin{eqnarray}
    \label{eq.excahnge-single-layer}
    \mathcal{U}_{J} = J \sum_{<ij>}\mathbf{S}_i\cdot\mathbf{S}_j = \frac{3 J}{2} S^2 (\bm{\alpha}\cdot\bm{\alpha} + 2\beta_0^2).
\end{eqnarray}
As anticipated the hard modes are penalized by the exchange. Similarly, we can expand the DM interaction and the easy axis anisotropy:
\begin{equation}
    \label{eq.DM-single-layer}
    \mathcal{U}_{DM} = \frac{D\sqrt{3}}{2}S^2 \left[3\bm{\alpha}\cdot\bm{\alpha} + 2(\beta_0^2 + \bm{\beta}\cdot\bm{\beta}) \right].
\end{equation}
The DM interaction in this approximation acts as a combination of an easy plane anisotropy and nearest neighbour exchange. As a result of the easy plane nature the $\bm{\beta}$ modes are no longer soft. They are however still degenerate as the DM interaction does not distinguish an easy direction, the modes retain an $O(2)$ symmetry.
\begin{eqnarray}
    \label{eq.aniso-single-layer}
    \mathcal{U}_{\text{easy-axis}} &=& -\sqrt{\frac{3}{2}} S^2 \delta (\alpha_x\cos 2\psi + \alpha_y\sin 2\psi) \\ \nonumber
    &+& \frac{S^2\delta}{2} (\beta_0^2 + \bm{\beta}\cdot\bm{\beta}) \\ \nonumber
    &-& \frac{S^2\delta}{4}(2\alpha_x^2 - \alpha_y^2 + \beta_x^2 - \beta_y^2 + 2\sqrt{2}\beta_0\beta_y)\cos 2\psi \\ \nonumber
    &-& \frac{S^2\delta}{4}(4\alpha_x\alpha_y + 2 \beta_x\beta_y + 2\sqrt{2}\beta_0\beta_x)\sin 2\psi,
\end{eqnarray}
where $\psi = \alpha_0/\sqrt{3}$. It is clear that the easy axis interaction kills any remaining continuous symmetry in the kagome plane and splits the degeneracy between the $\beta_x$ and $\beta_y$ modes. We can rely on the hierarchy of the energy scales in Mn$_3$Ge and obtain a fairly robust effective field theory for the soft modes $(\alpha_0,\bm{\beta})$. The total energy to quadratic order in fields is given by
\begin{equation}
\label{eq.U-J-D-delta}
    \mathcal{U}_{JD\delta} (\alpha_0,\bm{\alpha},\beta_0,\bm{\beta}) =  \mathcal{U}_{J} +  \mathcal{U}_{DM} + \mathcal{U}_{\text{easy-axis}}.
\end{equation}
Using this energy density we can solve for the modes $\bm{\alpha}$ and $\beta_0$: $(\delta \mathcal{U}_{JD\delta}/\delta\alpha_x) = (\delta \mathcal{U}_{JD\delta}/\delta\alpha_y) = (\delta \mathcal{U}_{JD\delta}/\delta\beta_0) = 0$. The solution is in terms of the fields $(\bm{\beta},\alpha_0)$. Plugging the solutions back into the total energy density Eq.~(\ref{eq.U-J-D-delta}) we get the effective theory at the $\Gamma$ point.
\begin{eqnarray}
\label{eq.U-eff-J-D-delta}
    \mathcal{U}_{\text{eff}} &=& \left(\sqrt{3} D + \frac{\delta}{2}\right)\bm{\beta}\cdot\bm{\beta}  \\ \nonumber
    &-& \frac{\delta}{4} \left[ (\beta_x^2 - \beta_y^2) \cos 2\psi - 2\beta_x\beta_y \sin 2\psi \right]  - \frac{\delta^3}{4 J^2} \cos 6\psi,
\end{eqnarray}
where we have used $J\gg D \gg \delta$ to retain the leading terms. Note that on integrating out the hard modes $\bm{\alpha}$, the easy axis anisotropy produces a six-fold symmetric anisotropy for the ground state expressed through $\psi$. These six states sit on the even hour marks of a 12 state clock for Mn$_3$Ge, see Fig.~\ref{fig.geometry-modes}. 

To this picture we now add two forms of strain---a unixaial strain in the kagome plane represented by a strain vector $(\epsilon_{xx} - \epsilon_{yy} , 2\epsilon_{xy} )$ and one that involves one axis in the kagome plane and the c-axis $(\epsilon_{xz},\epsilon_{yz})$. This will produce two effective theories which we then use to determine modifications to the anomalous Hall effect and the energetics of the spin waves.

\subsection{Strain}

Uniaxial strain in these group of kagome antiferromagnets Mn$_3$X (X = Ge, Sn, Ga, ...) produces an extra easy axis anisotropy which can be used to control the magnetic ground state. The strains we are going to work with are of two types---in plane shear $\epsilon_{xx} - \epsilon_{yy}$, $\epsilon_{xy}$ and out of plane shear $\epsilon_{xz}$, $\epsilon_{yz}$. Strain is represented by a symmetric rank-3 tensor $\epsilon_{ij} = (\partial_i u_j + \partial_j u_i)$ with $\mathbf{u}$ as the displacement field on the lattice. The effect of strain in the Hamiltonian in Eq.~(\ref{eq.minimal-H}) is captured through the variation of the Heisenberg exchange with lattice site displacements $\sum_{ij} \left[ (\partial J / \partial \mathbf{u}) \cdot \delta\mathbf{u}\right]\mathbf{S}_i\cdot\mathbf{S}_j$, following \cite{tchernyshyov2002-prl,tchernyshyov2002,dasgupta2021,ikhlas-future}.

The exact form of the decay of $J_{ij}$ with separation is not important and we retain only the first derivative correction. This correction is substantial in Mn$_3$X as evident from the very strong magnon phonon coupling in the antichiral 120$^\circ$ phase seen and calculated in \cite{Chen-mn3ge-2020} and also estimated in \cite{sukhanov2018} through measurement of magnetic order under pressure. Variations in the other energy scales are assumed to be much smaller and are neglected.

Further, evidence of a large response to strain is presented in  \cite{wang2019integration,guo2020giant}, where epitaxial strains are used to effect large changes in the anomalous Hall responses of  Mn$_3$Sn and Mn$_3$Ga respectively. We show that uniaxial strains strongly modify the anomalous Hall signal and the induced magnetization through the assumed coupling. An exact estimate of $\partial J/\partial \mathbf{u}$ in our case will need further microscopic modeling. Here we absorb this magnitude into strain.

The in plane strain produces a distortion of the 120$^\circ$ ground state and induces a net strain dependent moment. This is reflected in the energy density:
\begin{eqnarray}
\label{eq.energy-strain-1-single-layer}  \nonumber
    \frac{\mathcal{U}_{s1}}{3 S^2} &=& \frac{\epsilon}{8} (2\sqrt{6}\alpha_x - \alpha_x^2 + \alpha_y^2 + \beta_x^2 - \beta_y^2 - 2\sqrt{2}\beta_0\beta_y )\cos 2\psi_\epsilon \\ 
    &-& \frac{\epsilon}{8} (2\sqrt{6}\alpha_y + 2 \alpha_x\alpha_y - 2\beta_x\beta_y + 2\sqrt{2}\beta_0\beta_x ) \sin 2\psi_\epsilon.
\end{eqnarray}
Integrating out the same modes as before, $\alpha_x$, $\alpha_y$ and $\beta_0$ now gives us the effective theory as:
\begin{eqnarray}
\label{eq.u-eff-strain-in} \nonumber
    \mathcal{U} &=& \mathcal{U}_{eff} - \frac{3\delta\epsilon}{2J}\cos 2(\psi + \psi_\epsilon) \\ 
    &+& \frac{3 \epsilon}{8} \left[ (\beta_x^2 - \beta_y^2)\cos 2\psi_\epsilon + 2\beta_x\beta_y\sin 2\psi_\epsilon \right] 
\end{eqnarray}
where $\mathcal{U}_{eff}$ is shown in Eq.~(\ref{eq.U-eff-J-D-delta}). The shear strain introduces a second set of easy axes in the kagome plane as expected. Looking at the first term, we can see that this induced easy axes fights with the six fold axes from the local anisotropy. This provides the leverage to shear engineer the ground state and hence the transport properties of the sample. The secondary effect is through the effect on the $\bm{\beta}$ manifold \cite{ikhlas-future}. Here, the shear strain further splits the degeneracy between the two $\beta$ modes but now in an adjustable manner. 

The interlayer strain potential is given by variations of the interlayer exchanges with moment locations. Here we consider the variation of the nearest interplanar exchange $J_1$ \cite{Chen-mn3ge-2020}. To linear order in the normal modes this choice does not matter as the form of the coupling is dictated by symmetry. The higher orders will be affected by which exchange $J_i$ we examine.

The strain can be represented by the vector $(\epsilon_{xz},\epsilon_{yz}) = \epsilon_z(\sin \psi_{\epsilon z},\cos \psi_{\epsilon z}) = \bm{\epsilon}_z$. This couples to the magnetization vector formed from distortion of the ground state in the kagome plane $\mathbf{m} = (\alpha_y, - \alpha_x)$ to give an energy density---$\bm{\epsilon}_z\cdot\mathbf{m}$. This is the linear coupling between interplanar shear strain and the local magnetic order. Considering $J_1$ to be the exchange whose variation is strongest we can get the coupling to quadratic order in fields as:
\begin{eqnarray} \nonumber
\label{eq.energy-strain-2-single-layer}
    \frac{\mathcal{U}_{s2}}{3S^2}  &=&  \sqrt{\frac{3}{2}}(\alpha_y\epsilon_{xz} - \alpha_x\epsilon_{yz}) + \frac{\sqrt{3}\epsilon_{zz}}{4}(\bm{\alpha}^2 + 2\beta_0^2) \\ \nonumber
    &+& \frac{\epsilon_{xz}}{4}(2\alpha_x\alpha_y - 2\beta_x\beta_y + 2\sqrt{2}\beta_0\beta_x) \\
    &+& \frac{\epsilon_{yz}}{4}(\alpha_x^2 - \alpha_y^2 - \beta_x^2 + \beta_y^2 + 2\sqrt{2}\beta_0\beta_y).
\end{eqnarray}
This can be then used to obtain the effective theory with an out of plane shear strain. While writing this down we assume that the interplanar strain is much smaller than any intra kagome layer strain. We retain terms to leading order in the strain.
\begin{eqnarray}
     \mathcal{U} &=& \mathcal{U}_{eff} + \frac{3}{2} \frac{\delta \epsilon_z}{J} \sin(2\psi - \psi_{\epsilon z}) \\ \nonumber
     &-& \frac{3 \epsilon_z}{4} \left[ (\beta_x^2 - \beta_y^2)\sin(\psi_{\epsilon z}) + 2\beta_x\beta_y \cos(\psi_{\epsilon z}) \right].
\end{eqnarray}
This too affects the splitting of the $\beta$ modes like the interplanar strain. Note that unlike the interplanar strain the first term which couples the local anisotropy to the strain is linear in the small angle difference between the two sets of easy axes. This changes the switching response of the Hall signal we had previously obtained for the interplanar strain.

\section{Hall Response}

The modern theory of the anomalous Hall effect \cite{Karplus:1954, Sundaram:1999, Jungwirth:2002, Nagaosa:2010} relates the ``intrinsic'' part of the Hall conductivity tensor $\sigma^H_{ij}$ to the Hall vector $\mathbf{Q}$ representing a fictitious magnetic field experienced by electrons in momentum space:
\begin{equation}
\sigma^H_{ij} = \frac{e^2}{2\pi h} \epsilon_{ijk} Q_k,
\label{eq:K-def}
\end{equation}
For Mn$_3$X, $K_z=0$ as $\sigma^H_{xy} = 0$ \cite{Nakatsuji2015}. Thus the Hall vector can serve as an order parameter for the spontaneously broken symmetry of global rotations in the $xy$ plane. In \cite{Liu:2017} they show that to quadratic order the Hall vector $\mathbf{Q} \propto \mathbf{K} =  (\cos\psi,\sin\psi,0)$, with $\sqrt{3}\psi = \alpha_0$. The constant of proportionality is given by the electronic structure, which they calculate using the Kubo formalism. Thus the \textit{magnetic} normal mode $\alpha_0$ corresponding to uniform rotations of all three spins of the sublattice \cite{dasgupta-mn3ge-2020} is linearly linked to the Hall vector and any manipulations of the magnetic order 120$^\circ$ in the kagome plane will modify the Hall conductivity. The $\alpha_0$ mode in Mn$_3$X is penalized by the local anisotropy ($\delta$) which leads to a six-fold degenerate ground state \cite{Chen-mn3ge-2020,dasgupta-mn3ge-2020,Liu:2017}. These can be represented on a clock as shown in Fig.~\ref{fig.geometry-modes} with odd (even) hours representing states of Mn$_3$Ge (Sn). An external probe can be used to switch between the states on this manifold \cite{Takeuchi2021}. In our case we exercise this through shear strains, in plane---$(\epsilon_{xx}-\epsilon_{yy},2\epsilon_{xy}) = \epsilon (\cos 2\psi_\epsilon,\sin 2\psi_\epsilon)$, and out of plane---$(\epsilon_{xz},\epsilon_{yz}) = \epsilon_z (\cos \psi_{\epsilon z},\sin \psi_{\epsilon z})$.

The energy barriers separating these six grounds states is $\simeq \delta\sqrt{\delta/J}$ and is tiny as $\delta \ll J$ \cite{Chen-mn3ge-2020}. Thus we can use a very small strain ($\epsilon \equiv \delta \simeq .1 \% J$) to cause this switching, in Mn$_3$Sn we found this to be around $120$ MPa \cite{ikhlas-future}. The other compounds in the group are expected to have switching strains of similar magnitude and can be easily applied through epitaxial mismatch for instance \cite{guo2020giant,wang2019integration}. In our calculations we measure all energies in units of exchange $J$ which we set to one. This can be appropriately adjusted to specific compounds.
\begin{figure}[t]
		\centering
		\includegraphics[width =.9\columnwidth]{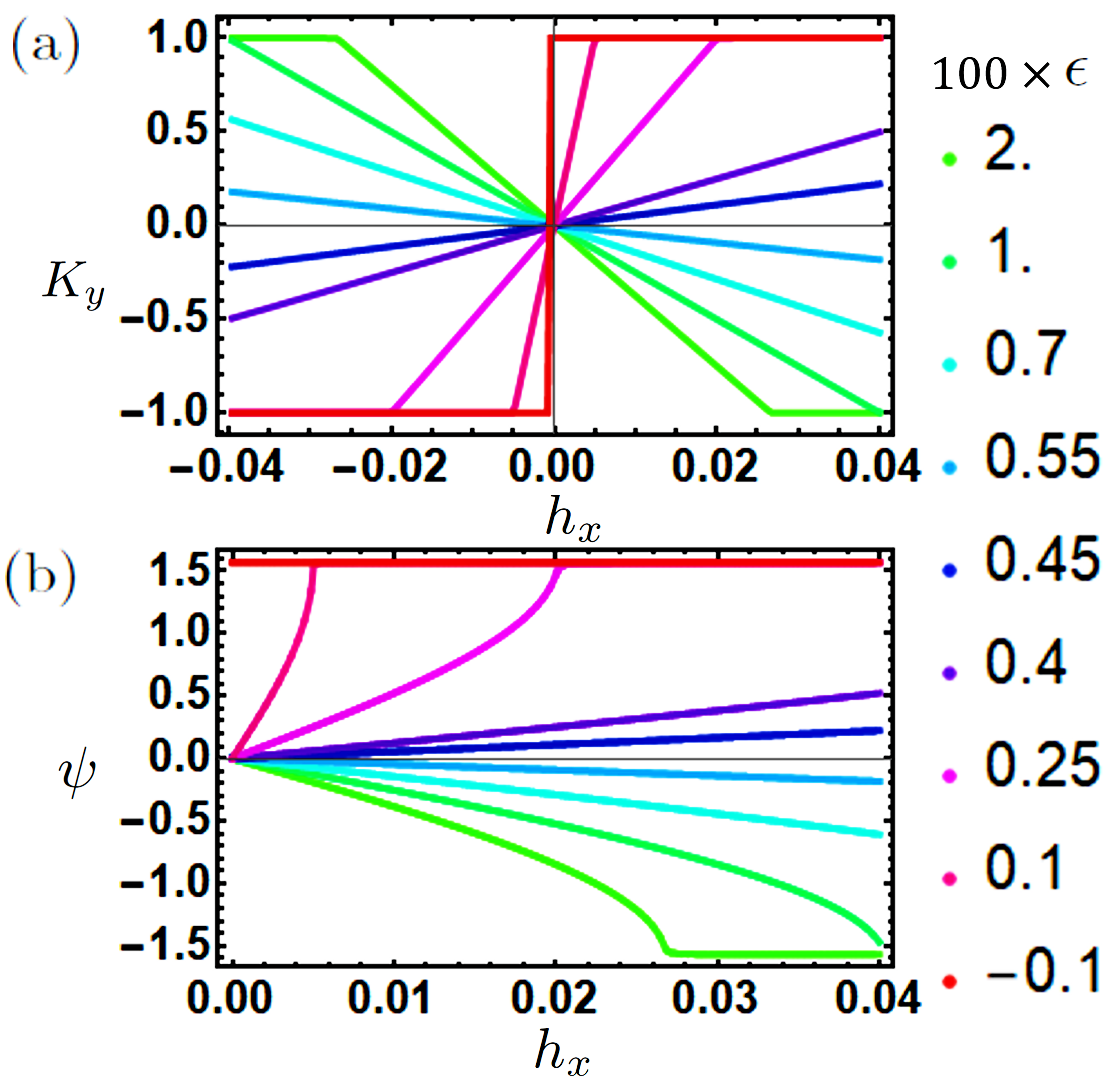}
		\caption{In (a) $K_y$ is shown with varying magnetic field and strength at $\epsilon_z = 0$, $\psi_\epsilon = \pi/2$ and $\psi_h = 0$ obtained from Eq.~(\ref{eq.reduced-f-plane-strain}). We compare all our energy scales to the exchange energy, the strain on the system $\epsilon$ is shown in the color shades with the value on the panel representing the ratio $100 \times \epsilon$. The anisotropy is set to $\delta = 0.005~J$. Panel (b) shows the variation of the angle $\psi$ with strain at positive fields. We can see a clear flip in the sign of the Hall vector as strain crosses $\epsilon_c = \delta$.}
		\label{fig.Hall-1}
\end{figure}
The primary difference between the Sn and Ge compounds is the reversal of the local easy axis direction which will change the critical strain at which we can flip the Hall response for a fixed magnetic field. Besides outlining this modified flipping strain, we will also analyze the situation where the strain is applied in three dimensions. Since the Hall response is confined to the kagome plane we can restrict our effective theory to the $\alpha$ modes.

The switching is a linear effect in strain and can be captured in a theory to linear order in the normal modes. In addition to the exchange, strain, and local anisotropies we introduce an external magnetic field $\mathbf{h} = h(\cos\psi_h,\sin\psi_h,0)$. The energy density calculated from Eq.~(\ref{eq.minimal-H}) in terms of the $\alpha$ modes:
\begin{eqnarray} \nonumber
\label{eq.Hall-strain-kagome}
     \mathcal{U}_{Hall} &=& \frac{3 J}{2} (\bm{\alpha}\cdot\bm{\alpha}) - \sqrt{\frac{3}{2}}\delta (\alpha_x\cos 2\psi + \alpha_y\sin 2\psi) \\ \nonumber
     &-& \sqrt{\frac{3}{2}} h [\alpha_y \cos(\psi - \psi_h) - \alpha_x\sin(\psi -\psi_h)] \\ \nonumber
     &-& \frac{3}{2}\sqrt{\frac{3}{2}} \epsilon (\alpha_y\sin 2\psi_\epsilon - \alpha_x\cos 2\psi_\epsilon) \\ 
     &+& 3 \sqrt{\frac{3}{2}} \epsilon_z (\alpha_y\cos \psi_{\epsilon z} - \alpha_x\sin \psi_{\epsilon z}). 
\end{eqnarray}
The derivation of these terms in presented in \cite{dasgupta-mn3ge-2020}. The last two terms here are new and are derived from the strain correction to the exchange, $\sum_{ij} \left[ (\partial J / \partial \mathbf{u}) \cdot \delta\mathbf{u}\right]\mathbf{S}_i\cdot\mathbf{S}_j$ expressed in terms of the $\bm{\alpha}$ modes.

Note that these terms effectively introduce a new set of easy axes in the kagome plane. From this we can integrate out the $\bm{\alpha}$ doublet using their equations of motion. This leaves a theory in terms of the azimuthal field $\psi$. In all future expressions we set $J=1$ to simplify our expressions, the factors of $J$ can be reinserted through dimensional analysis. We start with the in plane shear strain and set $\epsilon_z = 0$. This situation has been studied in detail for Mn$_3$Sn in \cite{ikhlas-future}. The resulting functional is:
\begin{eqnarray}
\label{eq.U-kagome-plane-strain-psi}
    \mathcal{F}_{\epsilon}  &=& - h \delta \sin(\psi + \psi_h) +  \delta\epsilon \cos [2(\psi + \psi_\epsilon)] \\ \nonumber
    &-& h\epsilon\sin(\psi - \psi_h + 2\psi_\epsilon),
\end{eqnarray}
where we have redefined the strain amplitude by $3/2$ to get a simpler expression. In experiments, the data for the critical strain is obtained in relation to the anisotropy energy scale ($\delta$) and this redefinition does not show up. We will analyze a simpler situation with $\psi_h = 0$ and $\psi_{\epsilon} = \pi/2$ as done in \cite{ikhlas-future} for Mn$_3$Sn. This gives an energy functional in terms of $\sin\psi = K_y$:
\begin{equation}
\label{eq.reduced-f-plane-strain}
    f_{\epsilon} = h(\epsilon - \delta)K_y + 2\delta\epsilon K_y^2.
\end{equation}
This is already a remarkable result, the Hall vector direction is perpendicular to the applied field. This is not unexpected in these compounds as the pseudotensor order parameter which is identified with the Hall vector does not follow the magnetic field. The three Hall phases isolated in Mn$_3$Sn \cite{ikhlas-future} can be identified here as well. Let us fix $\delta > 0$ and vary strain from $\epsilon > \delta$ to $\epsilon < 0$. 

For $\epsilon > 0$ the parabola has a positive quadratic part and hence its extremum is a minimum of $f_{\epsilon}$ given by $K_y = h(\delta - \epsilon)/4\delta\epsilon$. If we consider $h>0$ then for $\epsilon > \delta$, $\sgn (K_y) = -\sgn(h)$. This sign switches at $\epsilon_c = \delta$.  The slope of the Hall vector with field strength $dK_y/dh = (\delta - \epsilon)/4\delta\epsilon$ is positive for $\epsilon < \delta$ and negative otherwise. The negative slope identifies the diahallic regime, while the positive slope identifies the parahallic regime. For sufficiently large field, $h$, the Hall vector saturates at $h_s = |4\delta\epsilon/(\delta - \epsilon)|$. For $\epsilon < 0$ the quadratic part is negative and hence the functional has a maximum, implying that $|K_y| = 1$ with the sign decided by the magnetic field, $K_y = \sgn h$.

This demarcates a ferrohallic region shown by the red dotted line in Fig.~\ref{fig.Hall-1}. Notice the sharp transition in $K_y$ with the sign of the magnetic field, this is because here $K_y$ is independent of the magnitude of the magnetic field unlike the diahallic and parahallic regimes discussed before. The critical strain obtained for Mn$_3$Sn was $120$ MPa \cite{ikhlas-future}, given that this is equal to the anisotropy $\delta$, we expect this value to be halved in Mn$_3$Ge where the anisotropy is halved ($.001$ meV) \cite{Chen-mn3ge-2020}.

Let us now consider the case of the interplanar strain applied in isolation. The magnitude of this is modulated by variations of the interplanar exchanges with displacements in the location of the spin sites. The energy functional, Eq.~(\ref{eq.Hall-strain-kagome}) is then minimized with respect to the $\bm{\alpha}$ doublet setting $\epsilon = 0$. This gives us a theory for the Hall vector $\mathbf{K}$ and the strain field as:
\begin{eqnarray}
\label{eq.U-interplanar-strain-psi}
    \mathcal{F}_{\epsilon z}  &=& - h \delta \sin(\psi + \psi_h)  +  \delta\epsilon_z \sin (2\psi - \psi_{\epsilon z}) \\ \nonumber
    &+& h\epsilon_z\cos(\psi - \psi_h - \psi_{\epsilon z}),
\end{eqnarray}
where here we have absorbed a factor of $3$ into the definition of $\epsilon_z$. This functional is generically different from the in plane strain shown in Eq.~(\ref{eq.U-kagome-plane-strain-psi}). However, if we look at the same configuration: $\psi_h = 0$ and $\psi_{\epsilon z} = \pi/2$, the reduced functional:
\begin{equation}
\label{eq.reduced-f-interplane-strain}
    f_{\epsilon z} = h(\epsilon_z - \delta)K_y + 2\delta\epsilon_z K_y^2,
\end{equation}
identical to the expression obtained for the in-plane strain and we will get the same phases as we tune the strain $\epsilon_z$. The switch from ferro to diahallic behaviour happens at $\epsilon_c = \delta$ as before. 
In most experimental situations where we have a multilayer samples the two forms of strain will be combined. This can potentially lead to a complicated situation where we now have two external energy scales $\epsilon$ and $\epsilon_z$ competing with the local anisotropy $\delta$. The field theory in this case is:
\begin{eqnarray}
\mathcal{F}_{\text{strain}} &=& - h\delta \sin(\psi + \psi_h) + \epsilon \delta \cos[2(\psi + \psi_\epsilon)]  \\ \nonumber
&-& \epsilon h \sin(\psi - \psi_h + 2\psi_{\epsilon}) \\ \nonumber
&+& \epsilon_z \delta \sin(2\psi - \psi_{\epsilon z}) + \epsilon_z h \cos(\psi - \psi_h -\psi_{\epsilon z}).
\end{eqnarray}
This simplifies to:
\begin{equation}
    f_{strain} = h(\epsilon + \epsilon_z - \delta)K_y + 2\delta (\epsilon + \epsilon_z)K_y^2,
\end{equation}
with $\psi_h = 0$ and $\psi_\epsilon = \psi_{\epsilon z} = \pi/2$. From this expression it is clear that in this configuration the two strains are additive in their effect and cannot be disentangled. Say $\varepsilon = \epsilon + \epsilon_z$, if we start with $\varepsilon < 0$ we are in the ferrohallic regime and the diahallic regime for $\varepsilon > 0$. The sign switch is at $\varepsilon = \delta$.

\subsection{Pure strains}
Even in the absence of a magnetic field, $h = 0$, the Hall signal can be manipulated by strain. The functional in that case:
\begin{equation}
    \mathcal{F}_{\text{strain}}\vert_{h = 0} = \delta[\epsilon\cos(2\psi + 2\psi_\epsilon) + \epsilon_z\sin(2\psi + \psi_{\epsilon z})].
\end{equation}
The minima of this expression is controlled by the relative signs and magnitudes of $\epsilon$ and $\epsilon_z$. Consider $\sgn(\epsilon) = \sgn(\epsilon_z)$ and $\psi_\epsilon = \psi_{\epsilon z} = \pi/2$ as before. Then the Hall angle takes the value of $\psi = 0$ for $\sgn(\epsilon)>0$ and $\psi = \pi/2$ for $\sgn(\epsilon) < 0$. These two cases represent $K_y = 0$ and $K_y = 1$. Alternatively, we can apply two planar strains of different magnitudes $\epsilon_1$ and $\epsilon_2$ at $\pi/4$ to each other, $\psi_{\epsilon-1} = \psi_{\epsilon}$ and $\psi_{\epsilon-2} = \psi_{\epsilon} + \pi/4$. This produces an effective strain:
\begin{equation}
    \bm{\epsilon} = \sqrt{\epsilon_1^2 + \epsilon_2^2}~[\cos2(\psi_\epsilon + \zeta), \sin 2(\psi_\epsilon + \zeta)],
\end{equation}
where $2\zeta = \arctan[\epsilon_2/\epsilon_1]$. We can now vary the effective strain axis by changing the relative strength $\epsilon_1/\epsilon_2$ and vary the angle $\psi$ and hence the Hall response that way.

\begin{figure}[t]
		\centering
		\includegraphics[width =.9 \columnwidth]{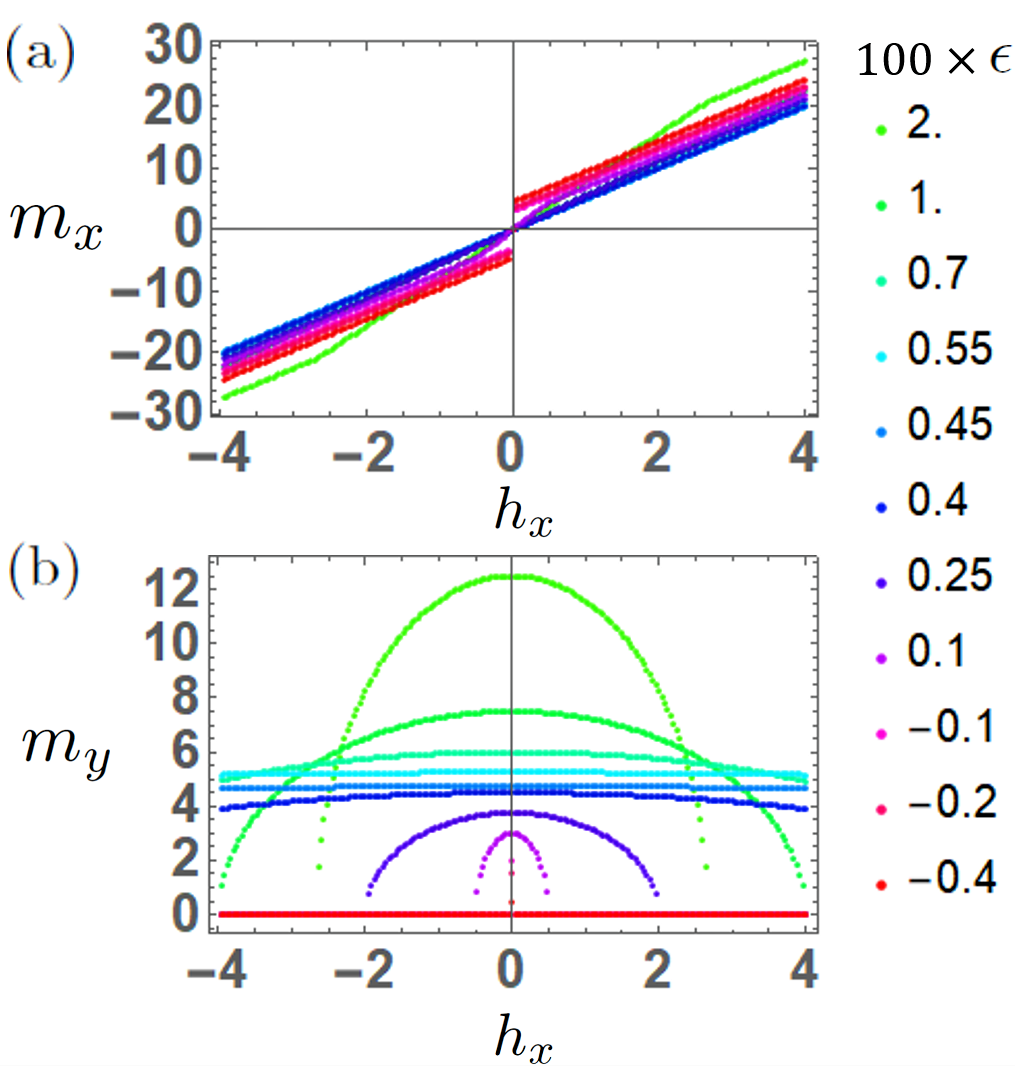}
		\caption{The variation of the magnetization along the external magnetic field $m_x$ (a) and perpendicular to it $m_y$ (b) for the setup shown in Fig.~\ref{fig.Hall-1}. The magnetization is given in units of milli$-\mu_B$. As detailed in the text, unlike the Hall vector magnetization does not reverse its sign with application of strain. Noticeably, the simulations show that $m_y$ for small magnetic fields is almost as large as the magnetization along the field.}
		\label{fig.magnetization}
\end{figure}

\section{Magnetization}

\subsection{In plane magnetization}
Just as it affects the Hall vector by introducing a secondary set of easy axes, uniaxial strain also induces a magnetic moment in the process. The induced magnetic moments can be represented by the $\bm{\alpha}$ doublet:
\begin{eqnarray}
m_x &=& \gamma S \sqrt{\frac{3}{2}} (\alpha_y \cos\psi - \alpha_x\sin\psi) \\ \nonumber
m_y &=& \gamma S \sqrt{\frac{3}{2}} (\alpha_x \cos\psi + \alpha_y\sin\psi),
\end{eqnarray}
where $\gamma$ is the gyromagnetic ratio converting the spin to a magnetic moment $(\mathbf{m} = \gamma\mathbf{S})$. We can plug into these the solutions for $\bm{\alpha}$ obtained from minimizing $\mathcal{U}_{Hall}$, Eq.~(\ref{eq.Hall-strain-kagome}). With $S = 1$, the magnetic moments in terms of the order parameter angle $\psi$ and the strain angles $\psi_\epsilon$ and $\psi_{\epsilon z}$.
\begin{eqnarray}
m_x &=& \gamma \frac{\gamma h \cos\psi_h + \delta \sin\psi}{2} \\ \nonumber
&+& \gamma\frac{\epsilon \sin(\psi + 2\psi_\epsilon) - \epsilon_z \cos(\psi - \psi_{\epsilon z})}{2}. \\ \nonumber
m_y &=& \gamma \frac{\gamma h \sin\psi_h + \delta \cos\psi}{2} \\ \nonumber
&-& \gamma\frac{\epsilon \cos(\psi + 2\psi_\epsilon) + \epsilon_z \sin(\psi - \psi_{\epsilon z})}{2}.
\end{eqnarray}
The first part of the expressions are the moments induced by the magnetic field $\mathbf{h}$ and the local anisotropy. The latter terms are the contribution from strain. Note that these are linear in strain---indicative of piezomagnetism. Using the same setup as the Hall effect, $\epsilon_z = 0$, $\psi_h = 0$ and $\psi_{\epsilon} = \pi/2$, these expressions can be simplified:
\begin{eqnarray}
\label{eq.magnetization}
m_x &=& \frac{\gamma}{2}[\gamma h \cos\psi_h + (\delta - \epsilon )K_y] \\ \nonumber
m_y &=& \frac{\gamma}{2}[\gamma h \sin\psi_h + (\delta + \epsilon)K_x],
\end{eqnarray}
see Fig.~\ref{fig.magnetization}. To simplify the figure we have used Bohr-magneton units.

It is clear from the expressions that the strain contributions can be significant when $K_y$ or $K_x$ is near $\pm1$. Noticeably, the Hall vector and the induced moment are no longer parallel to each other and are in fact perpendicular when $K_y = \pm 1$, the piezomagnetic contribution is highest to $m_x$ and not $m_y$. While the Zeeman term ensures that there is a positive projection of the induced magnetic moment along the direction of the external field, the perpendicular component is not identically zero. Particularly when $K_y$ is lowered from $\pm 1$ and $|K_x| > 0$ this perpendicular contribution is finite and its magnitude depends on the strain applied, see Fig \ref{fig.magnetization}.

The critical strain $\epsilon = \delta$ is where the piezomagnetic contribution to $m_x$ vanishes and this can be independently used to fix the anisotropy $\delta$ in a magnetization measurement like in the case of Mn$_3$Sn \cite{ikhlas-future}. This quantity is poorly resolved in inelastic neutron data due to the extremely small $\alpha_0$ mode gap, and this will be a cleaner measure of the parameter. 

For $\psi_h = 0$ and $\psi_{\epsilon z} = \pi/2$, the expression for interplanar strain is identical to Eq.~(\ref{eq.magnetization}) with $\epsilon \to \epsilon_z$ and $\psi_\epsilon \to \psi_{\epsilon z}$. Note that unlike the Hall vector the magnetization along the external field does \textit{not} switch signs with strain---in the dia and parahallic regimes $K_y \propto (\delta - \epsilon)$ and hence the spontaneous component of the magnetization is proportional to $m_x \propto h(\delta - \epsilon)^2$. Thus the $\sgn(m_x) = \sgn(h)$ throughout the process. 

\subsection{Out of plane magnetization}
An out of plane magnetization will be expressed through a finite value for the $\beta_0$ mode. Fig.(\ref{fig.geometry-modes}). To energetically stabilize an out of plane canting we need to add a DM interaction with the DM vector lying in the kagome plane \cite{Liu:2017}. The term in the Hamiltonian is:
\begin{equation}
    \label{eq.DM2}
    H_{\text{DM2}} = \sum_{<ij>} D_{ij}\cdot(\mathbf{S}_i\times\mathbf{S}_j),
\end{equation}
where $D_{ij} = D_2 ~(\hat{\mathbf{z}}\times\hat{\mathbf{e}}_{ij}$) with $\hat{\mathbf{e}}_{ij}$ as the unit vector pointing from site $i$ to site $j$. In a first approximation we only consider the modes $(\alpha_0,\bm{\alpha},\beta_{0})$. The two soft $\beta$ modes are soft only in the pure Heisenberg exchange limit. They are energetically costly in the presence of DM interactions and we will take advantage of this to simplify our expressions. Expanding $H_{\text{DM2}}$ in terms of the normal modes leads to the energy functional:
\begin{eqnarray}
    \mathcal{U}_{\text{DM2}} &=& \sqrt{\frac{3}{2}}D_2S^2\beta_0 (\alpha_y\cos\psi_k - \alpha_x\sin\psi_k) \\ \nonumber
    &=& \sqrt{\frac{3}{2}}D_2S^2\beta_0 (\mathbf{K}\times\bm{\alpha}).
\end{eqnarray}
The net out of plane spin per plaquette is given by:
\begin{equation}
    m_z = \frac{S D_2}{8\sqrt{3}(J+J_1)^{2}} \left[ 2\delta\sin(3\psi) + 3\epsilon\sin(\psi + 2\psi_\epsilon)\right].
\end{equation}
Here it might be useful to recall that the net spin induced in the kagome plane by the local anisotropy $\delta$ in the absence of strain was $m = (S\delta/2J) \mathbf{K}$. Now, for $\epsilon = 0$ the net spin induced out of plane is proportional to $D_2 \delta/(J + J_1)^2$, reduced from the in plane component by a factor $D_2/(J+J_1)$. This makes it minuscule. Note that, unlike the in-plane component the out of plane component has a periodicity of $2\pi/3$. There is an additional contribution from the strain term which modifies both the size of the moment and its periodicity. We can in addition to this add a magnetic field of the form $\mathbf{h} = (h_p\cos\psi_h,h_p\sin\psi_h,h_z)$. In this case the induced out of plane moment is:
\begin{widetext}
\begin{equation}
    m_z = \frac{\gamma h_z}{2(J + J_1)}  + \frac{D_2}{8\sqrt{3}(J + J_1)^2} \left[ 2 S \delta \sin(3\psi) + 3 S \epsilon \sin(\psi + 2\psi\epsilon) + 2 \gamma h_p \sin(\psi_h + 2\psi)\right] .
\end{equation}
\end{widetext}

\section{Corrections to the Landau functional}

Now we consider the effect of introducing correction from two main sources. First we consider the effect of higher order (in soft modes) terms from the strain tensor $\delta$. Secondly, we consider the consequences of a misaligned straining direction and magnetic field in the kagome plane.

\subsection{Effects of anisotropy}
\noindent
Here we discuss the Landau functional where we retain higher order corrections from the local anisotropy $\delta$. This is done by retaining terms quadratic in the hard modes $\bm \alpha$ while expanding the local easy-axis interaction. We suppress the out of plane strain $(\epsilon_z = 0)$ as the corrections are of a similar nature. The minimization of the energy in this situation gives a Landau functional:

\begin{widetext}
\begin{eqnarray}
\label{eq.Landau_2ndorder}
   \mathcal{F}_\epsilon &=& \mathcal{F}_\epsilon[\mathcal{O}(1/J)]  + \frac{\delta}{12 J^2} \left[ S^2\delta^2\cos{(6\psi)} + h^2 \cos{(2\psi_h + 4\psi)} + S^2\epsilon^2 \cos{(4\psi_\epsilon + 2\psi)} \right] \\ \nonumber
    &+& \frac{S\delta}{6 J} \left[ h\delta \cos{(\psi_h + 5\psi)} + \delta\epsilon S \cos{(2\psi_\epsilon + 4\psi)}  +  h\epsilon \cos{(\psi_h + 2\psi_\epsilon + 3\psi)}  \right],
\end{eqnarray}
\end{widetext}

where the order $(1/J)$ functional is shown in Eq.~(\ref{eq.U-kagome-plane-strain-psi}). We can now see the six-fold anisotropy that rises from the local easy axis. The remaining terms will affect the exact value of strain at which the sign of the Hall vector reverses but these factors are suppressed by an extra power of the exchange strength (J) in the denominator which at least in the case of Mn$_3$Ge is a 100 times larger than any non-exchange energy scale, from fits to inelastic neutron data \cite{Chen-mn3ge-2020}. We can simplify this situation by again taking $\psi_h =  0$ and $\psi_{\epsilon} = \pi/2$. This produces a functional:

\begin{eqnarray}
   \mathcal{F}_\epsilon &=& \mathcal{F}_\epsilon[\mathcal{O}(1/J)] \\ \nonumber
   &+& \frac{\delta}{12 J^2} \left[ S^2\delta^2\cos{(6\psi)} + h^2 \cos(4\psi) + S^2 \epsilon^2 \cos(2\psi) \right] \\ \nonumber
   &+& \frac{S\delta}{6 J} \left[  h\delta \cos(5\psi) - \delta\epsilon S \cos{(4\psi)}  -  h\epsilon \cos{( 3\psi)}\right]
\end{eqnarray}
This expression can be now expressed in terms of $\mathbf{K}$ and will modify the variation of the Hall vector as strain is varied. These variations are however suppressed by an extra factor of $J$ as we state earlier.

\subsection{Misalignment of strain and magnetic field}
The other question had to do with a change in the reversal point stemming from the misalignment of the magnetic field and strain directions. In this case let us consider $\psi_\epsilon = pi/2 +\eta$ where $\eta \ll 1$ and $\psi_h = 0$. Plugging this into the first order Landau functional, Eq.~(\ref{eq.U-kagome-plane-strain-psi}):
\begin{eqnarray}
  \mathcal{F}_\epsilon (\eta) &=& \mathcal{F}_\epsilon + 2\eta \epsilon \cos\psi(h + 2\delta \sin\psi)\\ \nonumber
  &+& 2 \eta^2 \epsilon [\delta \cos(2\psi) - h\sin(\psi)]
\end{eqnarray}
To linear order in the correction angle, and assuming corrections to order $\eta K_y^2$, the energy functional reads:
\begin{equation}
    f_{\epsilon} = [h(\epsilon - \delta) + 4\eta\delta\epsilon]K_y + \epsilon(2\delta - h\eta) K_y^2.
\end{equation}
The condition for switching is now:
\begin{equation}
    \frac{h}{4\epsilon} - \frac{h}{4\delta} = \eta.
\end{equation}

\section{Spin wave dispersion for single layer}
We end our discussion of the effect of strain by showing that for a single kagome layer system, strain modification to the spin wave spectrum is not remarkable. This is primarily because the spin wave spectrum is dominated by the exchange energy terms which are orders of magnitude larger. We briefly recall that in the monolayer kagome system we had a spin wave theory of the form:
\begin{eqnarray}
 \mathcal{L} = \frac{\rho_\alpha}{2} \dot{\alpha}_0^2 + \frac{\rho_\beta}{2} \dot{\bm{\beta}}^2 - \frac{J S^2}{8 \sqrt{3}} [(\bm{\nabla}\alpha_0)^2 + 2(\bm{\nabla}\cdot\bm{\beta})^2],
\end{eqnarray}
where we have used the magnetic unit cell area $A = 2\sqrt{3}a^2$. The structure of the gradient term ensures that in its elasticity analogue the shear coefficient is zero which manifests as a dispersionless band \cite{dasgupta-mn3ge-2020}.

This theory is modified in the presence of strain. The first modification comes in the form of additional terms that are allowed. In these compounds the six fold symmetry allows terms that are $D_3$ symmetric. These terms can be constructed from three unit vectors $\mathbf{n}_1$, $\mathbf{n}_2$, and $\mathbf{n}_3$ which are aligned at 120$^\circ$ to one another. With these three unit vectors, one can form a sum:
\begin{equation}
    \mathcal{U} = \sum_{i = 1}^3(\mathbf{a}\cdot\mathbf{n}_i)(\mathbf{b}\cdot\mathbf{n}_i)(\mathbf{c}\cdot\mathbf{n}_i),
\end{equation}
where $\mathbf{a}$, $\mathbf{b}$, and $\mathbf{c}$ are three vectors in the $xy$ plane. This term is invariant under 120$^\circ$ rotations, squaring it gives a $D_3$ symmetric term. In the absence of strain the $\bm{\beta}$ doublet could form a term with $\mathbf{a} = \mathbf{b} = \bm{\nabla}$ and $\mathbf{c} = \bm{\beta}$. This contributes to the dispersion only at the quadratic order $\Delta \omega^2 \propto k^4$ \cite{dasgupta-mn3ge-2020}.

With strain in the mix, we have another vector $(\epsilon_{xx}-\epsilon_{yy}, 2\epsilon_{xy}) = \epsilon_0(\cos 2\psi_\epsilon,\sin 2\psi_\epsilon)$ in the $xy$ plane. The symmetry allowed term in this case is:
\begin{equation}
    \mathcal{U} = \epsilon_0^2 [(\partial_x\beta_y + \partial_y\beta_x)\cos 2\psi_\epsilon + (\partial_x\beta_x - \partial_y\beta_y)\sin 2\psi_\epsilon]^2.
\end{equation}
The correction to the dispersion now appears at the linear order $\Delta \omega^2 \propto k^2$. It is clear from the energy density that the correction is not isotropic, the expressions are simplified when $\psi_\epsilon = 0$ or $\psi_\epsilon = \pi/2$ as in most experimental setups:
\begin{eqnarray}
  \omega^2 &=& v_{||}^2 k^2 - \frac{\epsilon_0^2}{2\rho_\beta} k^2 \cos 2\phi (1-\cos 2\phi) \\ \nonumber
  \omega^2 &=& v_{\perp}^2 k^2 +  \frac{\epsilon_0^2}{2\rho_\beta} k^2 (2+\cos 2\phi) (1-\cos 2\phi),
\end{eqnarray}
where $\phi = \arctan(k_y/k_x)$ is the planar angle in k space.

In addition to this symmetry allowed term, strain also has a contribution to the dispersion at the linear order in $\epsilon$. At the linear order in fields strain shows up as a gauge field for the $\alpha_0$ singlet field. The singlet theory is modified:
\begin{equation}
    \mathcal{L}_{\alpha_0} = \frac{\rho_\alpha}{2}\dot{\alpha}_0^2 - \frac{A}{2}(\bm{\nabla}\alpha_0)^2 \to \frac{\rho_\alpha}{2}\dot{\alpha}_0^2 - (\mathbf{D}\alpha_0)^2,
\end{equation}
where $\mathbf{D} = \bm{\nabla} + \mathbf{a}$. Here $\mathbf{a}$ are strain dependent gauge fields $a_x = (3\epsilon_0/2J)(\cos 2\psi_\epsilon - \sqrt{3}\sin 2\psi_\epsilon)$ and $a_y = (3\epsilon_0/2J)(\sqrt{3}\cos 2\psi_\epsilon + \sin 2\psi_\epsilon)$. Since the strain is uniform in our setup this does not have an effect on the spin waves. However, if strain were to be a variable parameter then we would have a gauge field for the magnons. This is likely to happen at the sample edges where strain varies discontinuously.

At the quadratic order in fields strain contributes to both $\alpha_0$ and $\bm{\beta}$. The contribution to $\alpha_0$ can be ignored as there the modification to the spin wave velocity is minimal. The contribution to the $\bm{\beta}$ mode is more relevant since one of them is dispersionless.
\begin{figure}[]
		\centering
		\includegraphics[width = \columnwidth]{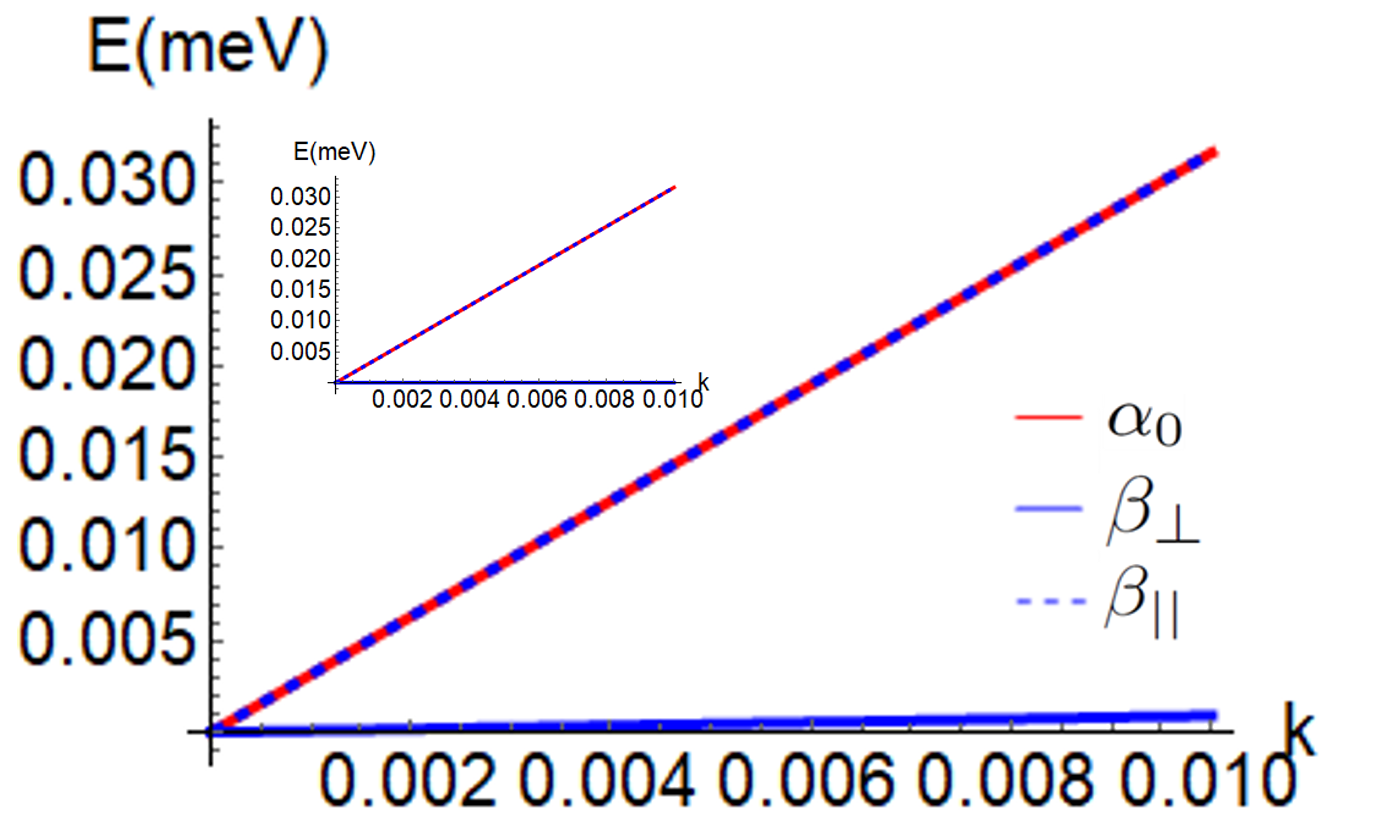}
		\caption{Dispersion with nearest neighbour interaction, $J$, and finite intralayer strain $\epsilon = J/100$ (zero strain in inset). The dispersion cut is along the line $\mathbf{q} = k(1,1)$. The strain lifts the flat $\beta_{\perp}$ mode. However, its effect on the degeneracy between the $\alpha_0$ and $\beta_{\perp}$ mode is only visible at fairly large values of the strain parameter.}
		\label{fig.dispersion}
\end{figure}

\section{Energy gaps}

We now calculate the modification of the spin wave gaps due to strain. We focus on the three modes which are not exchange penalized, these are the symmetric interlayer combinations: of $\alpha_0$ and $\bm{\beta}$. The antisymmetric fields are gapped by a substantial fourth nearest neighbour, interplanar, ferromagnetic Heisenberg exchange $J_4 \simeq 17$ meV \cite{Chen-mn3ge-2020}. The two inertias are given in terms of the exchange interactions \cite{dasgupta-mn3ge-2020}: 
\begin{equation}
    \rho_{\beta} = \frac{1}{3~(J_1 + J_2)} = 2\rho_{\alpha},
\end{equation}
where $(J_1 + J_2)$ is the effective nearest neighbor exchange. The in plane strain modifies the $\alpha_0$ energy gap as anticipated, since strain induces a uniaxial anisotropy. The correction is an order of magnitude $\mathcal{O}(\epsilon \delta/J)$, in exchange strength, stronger than just with the local anisotropy $\mathcal{O}(\delta^3/J^2)$.
\begin{equation}
   \rho_{\alpha} \omega^2_{\alpha_0} = \frac{\epsilon \delta}{J} (\cos 2\psi_\epsilon - \sqrt{3} \sin 2\psi_\epsilon) + 3 \left(\frac{\delta^3}{J^2}\right).
\end{equation}
Previously, we only had the second contribution \cite{Chen-mn3ge-2020,dasgupta-mn3ge-2020}. The uniaxial nature also splits the $\bm{\beta}$ bands: \begin{eqnarray}
 \rho_{\beta} \omega^2_{\beta_1} &=& 2\sqrt{3} D + \delta + \frac{3}{8}\epsilon(\cos 2\psi_\epsilon - \sqrt{3} \sin 2\psi_\epsilon) \\ \nonumber
 \rho_{\beta} \omega^2_{\beta_2} &=& 2\sqrt{3} D + \delta - \frac{3}{8}\epsilon(\cos 2\psi_\epsilon - \sqrt{3} \sin 2\psi_\epsilon).
\end{eqnarray}
Assuming that the DM interaction strength is greater than strain and the local anisotropy, $D \gg \epsilon_0,\delta$ we can get a compact form for the splitting of the bands:
\begin{equation}
    \Delta_{\beta} = \frac{3\epsilon}{8}\left(\frac{\cos 2\psi_\epsilon - \sqrt{3} \sin 2\psi_\epsilon}{\sqrt{\rho_\beta ( 2\sqrt{3} D + \delta)}}\right).
\end{equation}
Strain can then be used to control the dynamics of the $\alpha_0$ and $\bm{\beta}$ mode at a linear order. This maybe useful to engineer the response to externally applied spin torques, especially in the case of the $O(2)$ symmetric $\alpha_0$ mode which has a natural frequency scale of $\sqrt{\epsilon\delta/(\rho J)}$.

\section{Discussion}
In conclusion we have shown that both interplanar and intraplanar strain can be effectively used to engineer the Hall response of the non collinear antiferromagnet Mn$_3$Ge. In addition, we have shown that under strain the energy gaps at the $\Gamma$ point changes, showing that indeed strain creates a biaxial anisotropy. The resulting splitting of the $\bm{\beta}$ modes should be observable in neutron scattering and AFMR measurements. Recently, there has been a few works where the dynamics of the $\alpha_0$ mode has been directly probed in Mn$_3$Ge, namely in a Spin Transfer Torque setup \cite{Takeuchi2021} and a Josephson tunneling setup \cite{Jeon2021}. Strain will affect both these measurements since, as we show, it can be used to effectively modulate the $\alpha_0$ mode \cite{dasgupta2022tuning}. This makes the strain engineering of the Hall response a likely path to novel devices.

\begin{acknowledgments}
S.D. is supported by funding from the Max Planck-UBC-UTokyo Center for Quantum Materials, the Canada First Research Excellence Fund, Quantum Materials and Future Technologies Program, and the Japan Society for the Promotion of Science KAKENHI Grant No. JP19H01808. 
\end{acknowledgments}

\bibliographystyle{apsrev4-1}
\bibliography{main}
\end{document}